\documentclass[12pt,english]{article}
\usepackage{textcomp}
\usepackage[utf8]{inputenc}
\usepackage[english]{babel}
\usepackage{amsfonts,epsfig}
\newcommand{\be}{\begin{equation}}
\newcommand{\ee}{\end{equation}}

\newcommand{\beq}{\begin{eqnarray}}
\newcommand{\eeq}{\end{eqnarray}}
\newcommand{\nn}{\nonumber}
\newcommand{\bi}{\bibitem}
\large
\date{empty}

\begin{document}

%\preprint{APS/123-QED}

\title{\bf Calculation of the contribution to muon $g - 2$ due to the effective anomalous three boson interaction and the new experimental result}

\author{\bf B.A. Arbuzov and I.V. Zaitsev\\
Skobeltsyn Institute for Nuclear Physics of\\ Lomonosov Moscow State University}
%
%\email{arbuzov@theory.sinp.msu.ru}

\date{\today}% It is always \today, today,
             %  but any date may be explicitly specified

%\pacs{11.30.Rd; 12.38.Lg 12.39.-x; 12.40.Yx}% PACS, the Physics and Astronomy
                             % Classification Scheme.
%\keywords{Effective interaction; Compensation equation; QCD infrared region}%Use showkeys class option if keyword
                              %display desired
\maketitle

\begin{abstract}
Using the approach based on Bogoliubov compensation principle is applied to
calculation of a contribution to the muon $g-2$. Using the previous results on spontaneous generation of the effective anomalous three-boson interaction we
calculate the contribution, which proves to agree with the well-known discrepancy. The calculated quantity contains no adjusting parameters but the experimental values for the muon and the W-boson masses. The result can be considered as a confirmation of the approach.

\end{abstract}%\section{Compensation equation for anomalous tree-boson
%interaction}
Previous measurements of the anomalous magnetic moment of the muon $(g-2)_\mu = a_\mu$~\cite{g-2ex} provided a deviation of the experiment from predictions of the Standard Model. According to previous
analysis of the problem~\cite{g-2th2} this deviation $\Delta a_\mu$ safely comprise 3 standard deviations
\begin{eqnarray}
& &\Delta a_\mu\,=\,(276 \pm 80)\,10^{-11}\,;\label{amu}\\
& &\Delta a_\mu\,=\,(250 \pm 80)\,10^{-11}\,.\nonumber
\end{eqnarray}
Quite recent result reads \cite{g-3e}
\begin{equation}
\Delta a_\mu\,=\,(251 \pm 59)\,10^{-11}\,;\label{amunew}
\end{equation}
Deviations~(\ref{amu}) exceed zero a bit more than three SD, but the last result~(\ref{amunew}) may  be considered as the final establishing of the effect.
It should be emphasized, that the deviation from the SM calculations means
the deviation from perturbative calculations in the electro-weak theory. However there quite may be
non-perturbative contributions to physical quantities. In particular, the method of disclosing of the non-perturbative effects is developing starting of N.N. Bogoliubov compensation principle~\cite{NNB1, NNB2}.
In works~\cite{BAA04, BAA06, AVZ06, BAA07, AVZ09, BAA09,AZ11, AVZ2},
this principle
was applied to studies of a spontaneous generation of effective non-local interactions in renormalizable gauge theories of the Standard Model.

In the present letter we apply the previous results to the problem of the muon $\Delta a_\mu$. It will come clear, that the effect under discussion is quite natural in the theory with account of the spontaneous generation of an effective interaction in the conventional electro-weak theory.

The main principle of the approach is to check if an effective interaction
could be generated in the chosen variant of a renormalizable theory. In view
of this one performs "add and subtract" procedure for the effective
interaction with a form-factor. Then one assumes the presence of the
effective interaction in the interaction Lagrangian and the same term with
the opposite sign is assigned to the newly defined free Lagrangian.

In works~\cite{AZ11, AVZ2}  the approach was applied to the electro-weak interaction and a possibility of spontaneous generation of anomalous three-boson interaction of the form
\begin{equation}
-\,\frac{G}{3!}\cdot\,\epsilon_{abc}\,W_{\mu\nu}^a\,W_{\nu\rho}^b\,W_{\rho\mu}^c\,;
\label{FFF}
\end{equation}
was studied. In the present work we continue investigation
of the electro-weak theory using other approximation scheme,
which will be formulated in what follows.

The notation~(\ref{FFF}) means corresponding
non-local vertex in the momentum space
\begin{eqnarray}
& &(2\pi)^4\,G\,\,\epsilon_{abc}\,(g_{\mu\nu} (q_\rho pk - p_\rho qk)+ g_{\nu\rho}
(k_\mu pq - q_\mu pk)+\nonumber\\
& &g_{\rho\mu} (p_\nu qk - k_\nu pq)+ q_\mu k_\nu p_\rho - k_\mu p_\nu q_\rho)\times\nonumber\\
& &F(p,q,k)
\delta(p+q+k)+...;\label{vertex}
\end{eqnarray}
where $F(p,q,k)$ is a form-factor and
$p,\mu, a;\;q,\nu, b;\;k,\rho, c$ are respectfully incoming momenta,
Lorentz indices and weak isotopic indices
of $W$-bosons. We mean also that there are present four-boson, five-boson and
six-boson vertices according to the well-known non-linear expression for $W_{\mu\nu}^a$. Note, that in the approximation used we  maintain the gauge invariance of the approach.

Effective interaction~(\ref{FFF}) is
usually called anomalous three-boson interaction and it is considered for long time on phenomenological grounds~\cite{Hag}. Note, that the first attempt to obtain the anomalous three-boson interaction in the framework of Bogoliubov approach was done in work~\cite{Arb92}. Our interaction constant $G$ is connected with
conventional definitions in the following way
\begin{equation}
G\,=\,-\,\frac{g\,\lambda}{M_W^2}\,.\label{Glam}
\end{equation}
The current limitations for parameter $\lambda$ read~\cite{EW, EW13},
\begin{eqnarray}
& &\lambda\,=\,-\,0.016^{+0.021}_{-0.023}\,;\; -\,0.059< \lambda < 0.026\,
(95\%\,C.L.)\,.
\nonumber\\
& &\lambda_\gamma\,=\,-\,0.022\,\pm 0.019\,;\label{EW13}
\end{eqnarray}
where the last number~(\ref{EW13}) is obtained recently by joint analysis of LEP data by the four experimental groups: ALEPH, DELPHI, L3, OPAL.
We assume no difference in anomalous interaction for $Z$ and $\gamma$, i.e.
$\lambda_Z\,=\,\lambda_\gamma\,=\,\lambda$ according to standard relation
$W^0=\sin\theta_W A + \cos\theta_W Z$.

In works~\cite{AZ11, AVZ2}  the approach was applied to the electro-weak interaction and a possibility of spontaneous generation of anomalous three-boson interaction of the form~(\ref{FFF})
was studied. In the present work we continue investigation
of the electro-weak theory using other approximation scheme,
which will be formulated in what follows.

The goal of a study is a quest of an adequate
approach, the first non-perturbative approximation of
which describes the main features of the problem.
Improvement of a precision of results is to be achieved
by corrections to the initial first approximation.

This first approximation, corresponding to one-loop equation, is described in works~\cite{AZ11, AVZ2}, where we have studies the possibility of the existence of a non-trivial solution
with the following simple dependence on all three variables
\begin{equation}
F(p_1,\,p_2,\,p_3)\,=\,F(\frac{p_1^2\,+\,p_2^2\,+\,p_3^2}{2})\,;\label{123}
\end{equation}
Let us present the expression for four-boson vertex
\begin{eqnarray}
& &\frac{V(p,m,\lambda;\,q,n,\sigma;\,k,r,\tau;\,l,s,\pi)}{\imath\,(2 \pi)^4} =\nonumber\\
& &g G \biggl(\epsilon^{amn}
\epsilon^{ars}\Bigl(U(k,l;\sigma,\tau,\pi,\lambda)-U(k,l;\lambda,\tau,\pi,\sigma)-
U(l,k;\sigma,\pi,\tau,\lambda)+\nonumber\\
& &U(l,k;\lambda,\pi,\tau,\sigma)+U(p,q;\pi,\lambda,\sigma,\tau)-U(p,q;\tau,\lambda,\sigma,\pi)
-U(q,p;\pi,\sigma,\lambda,\tau)+\nonumber\\
& &U(q,p;\tau,\sigma,\lambda,\pi)\Bigr)-\epsilon^{arn}\,
\epsilon^{ams}\times\Bigl(U(p,l;\sigma,\lambda,\pi,\tau)-U(l,p;\sigma,\pi,\lambda,\tau)-
\nonumber\\
& &U(p,l;\tau,\lambda,\pi,\sigma)+U(l,p;\tau,\pi,\lambda,\sigma)+ U(k,q;\pi,\tau,\sigma,\lambda)-\label{four}\\
& &U(q,k;\pi,\sigma,\tau,\lambda)-U(k,q;\lambda,\tau,\sigma,\pi)+U(q,k;\lambda,\sigma,\tau,\pi)\Bigr)+
\nonumber\\
& &\epsilon^{asn}\,
\epsilon^{amr}\Bigl(U(k,p;\sigma,\lambda,\tau,\pi)-U(p,k;\sigma,\tau,\lambda,\pi)
+U(p,k;\pi,\tau,\lambda,\sigma)-\nonumber\\
& &U(k,p;\pi,\lambda,\tau,\sigma)-U(l,q;\lambda,\pi,\sigma,\tau)+
U(l,q;\tau,\pi,\sigma,\lambda)-\nonumber\\
& &U(q,l;\tau,\sigma,\pi,\lambda)+
U(q,l;\lambda,\sigma,\pi,\tau)\Bigr)\biggr);\nonumber\\
& &U(k,l;\sigma,\tau,\pi,\lambda)=(k_\sigma l_\tau g_{\pi\lambda}-k_\sigma\,
l_\lambda g_{\pi\tau}+\nonumber\\
& &k_\pi\,l_\lambda\,g_{\sigma\tau}-(kl)g_{\sigma\tau}g_{\pi\lambda}) F(k,l,-(k+l))\,.\nonumber
\end{eqnarray}
Here triad $p,\,m,\,\lambda$ {\it etc} means correspondingly momentum, isotopic index, Lorentz index of a boson.

Now in the way of studying the problem~\cite{AZ11, AVZ2}
we get convinced, that there exists a non-trivial solution of the compensation equation, which
has the the following form for $0 <z < z_0$\\
\begin{eqnarray}
& &F(z)\,=\,\frac{1}{2}\,G_{15}^{31}\Bigl( z\,|^0_{1,\,1/2,\,0,\,-1/2,\,-1}\Bigr) - \frac{85\,g \sqrt{N}}{512\,\pi}\,G_{15}^{31}\Bigl( z\,|^{1/2}_{1,\,1/2,
\,1/2,\,-1/2,\,-1}\Bigr)\,+\nonumber\\
& &C_1\,G_{04}^{10}\Bigl( z\,|1/2,\,1,\,-1/2,\,-1\Bigr)\,+
C_2\,G_{04}^{10}\Bigl( z\,|1,\,1/2,\,-1/2,\,-1\Bigr)\,.\label{solutiong}\\
& &z\,=\,\frac{N\,G^2\,x^2}{1024 \pi^2}\,;\quad x\,=\,p^2\,;
\nonumber
\end{eqnarray}
where
$$
G_{qp}^{nm}\Bigl( z\,|^{a_1,..., a_q}_{b_1,..., b_p}\Bigr)\,;
$$
is a Meijer function~\cite{BE}. In case $q=0$ we write only indices $b_i$ in one line. Constants $C_1,\,C_2$ are defined by boundary conditions.
For $ z > z_0$
\begin{equation}
F(z)\,=\,0\,.\label{F)}
\end{equation}

Parameters of solution~(\ref{solutiong}) are the following
\begin{eqnarray}
& &g\,=\,g(z_0)\,=\,0.60366\,;\quad z_0\,=\,9.61750\,;\nonumber\\
& &C_1\,=\,-\,0.035096\,;\quad
 C_2\,=\,-\,0.051104\,.\label{gY}
\end{eqnarray}

We would draw attention to the fixed value of parameter $z_0$. The solution
exists only for this value~(\ref{gY}) and it plays the role of eigenvalue.
As a matter of fact from the beginning the existence of such eigenvalue is
by no means evident. The definite value for $g(z_0)$  is also worth mentioning.

Emphasize, that an existence of a non-trivial solution of a compensation
equation is extremely
restrictive. In the most cases such solutions do not exist at all. When we start from a
renormalizable theory we have arbitrary value for
its coupling constant. Provided there exists {\it
stable non-trivial solution of a compensation equation} the
coupling is fixed as well as the parameters of this non-trivial solution.

\begin{figure}
\begin{center}
\includegraphics[width=8cm]{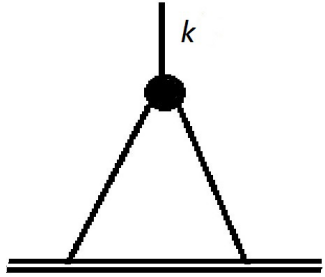}
\caption {One loop diagram for calculation of new contribution to the muon magnetic moment. Vertical line represents the photon, simple lines -- W bosons, black spot -- triple vertex~(\ref{FFF}) with corresponding form-factor. Double line represents the muon.}
\end{center}
\label{Fig1}
\end{figure}

\begin{figure}
\includegraphics[width=16cm]{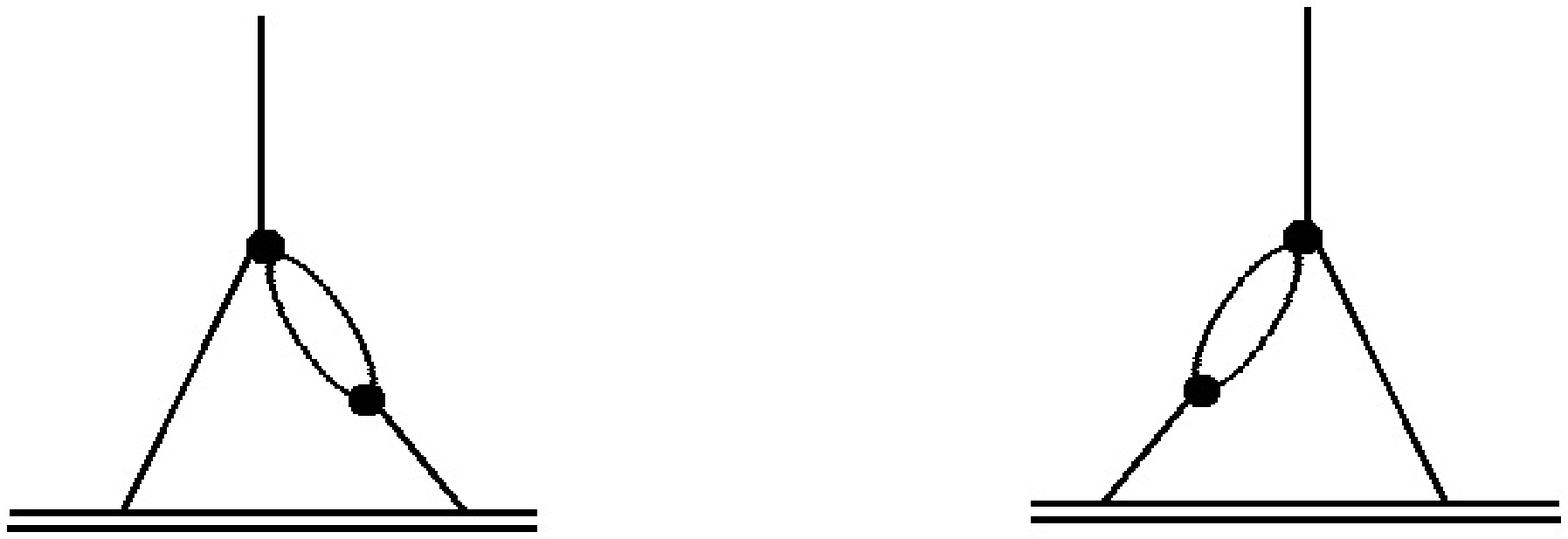}
\caption{Two loop diagrams for calculation of new contribution to the muon magnetic moment. Vertical line represents the photon, simple lines -- W bosons, black spots -- triple vertex~(\ref{FFF}) and four leg vertex~(\ref{four}) with corresponding form-factors. Double line represents the muon.}
\label{Fig2}
\end{figure}

Now let us consider a contribution of interaction~(\ref{FFF}) with form-factor defined by relations~(\ref{123},\ref{solutiong},\ref{gY}) to the anomalous magnetic moment of the charged spin one half particle. The first approximation described by the simplest diagram presented in Fig.\ref{Fig1} gives zero. However, the next approximation, presented by diagrams Fig.\ref{Fig2} leads to an interesting result. %\begin{center}
%\end{center}
The calculations are performed in the unitary gauge and give the following result after normalization of the gauge coupling $G$. In doing this we choose the part of the vertex~(\ref{four}), which gives contribution to usual gauge structure of triple vertex whereas the structure of the anomalous vertex~(\ref{FFF}) gives zero contribution to the magnetic moment.
\begin{eqnarray}
& &\Delta\,a\,=\,\frac{ m e g^2 G^2 N}{6 (16 \pi^2)^2 M_W^2}\nonumber\\
& &\int_0^Y dt F^2(t)\biggl(\int_0^t \frac{4 t y^2\,dy}{(6 t-3 y)(y+M_W^2)^2}+\nonumber\\
& &\int_t^{4 t/3}\frac{4 t y(16 t^3-48 t^2 y+48 t y^2-15 y^3)\,dy}{3(2 t-y)(y+M_W^2)^2}\biggr)\,.\label{da1}
\end{eqnarray}
From~(\ref{da1}) with definitions of variable $z$ and of the form-factor~(\ref{solutiong}, \ref{gY}) we obtain the following result for contribution to the magnetic moment
\begin{eqnarray}
& &\Delta a_\mu\,=\,\frac{g(z_0)^2\,m^2}{3\,\pi^2\,M_W^2}\,\biggl(20\,\ln\biggl[\frac{4}{3}\biggr]
-\frac{13}{3}\biggr)\times\nonumber\\
& &\int_0^{z_0}\, F^2(z)\,dz\,=\,2.775\times10^{-9}\,.\label{int2}
\end{eqnarray}
where we have used only values of the muon mass and the $W$-boson mass.
All other parameters are defined by solution~(\ref{solutiong}) with parameters~(\ref{gY}). Let us draw attention to the disappearance of the effective interaction coupling constant $G$ from expression~(\ref{int2}). This is due to entering of factor $G^2$
into the denominator according to definition of variable $z$. Thus the main result does not depend on $\lambda$. This parameter influence only the next approximations.
Let us estimate possible corrections due to
$M_W \neq 0$. They are defined by the following parameter
\begin{equation}
\frac{\sqrt{2} g |\lambda|}{32 \pi}\,=\,0.0005\,;
\end{equation}
with the maximal value of $|\lambda|=0.059$ from restrictions~(\ref{EW13}). Thus this correction may comprise only $0.05\%$.

We conclude that our calculations quite agree the last decisive measurement~(\ref{amunew})~\cite{g-3e}.
Note, that earlier our approach to the problem was discussed in work~\cite{A13} and in book~\cite{Abook}.
It also would be advisable to look for other effects of interaction~(\ref{FFF}) e.g.~\cite{3WA, 3WAZ}.

\end{document}